# Structure of Uniformly Rotating Stars

Robert G. Deupree

Institute for Computational Astrophysics, Department of Astronomy and Physics,
Saint Mary's University, Halifax, NS B3H 3C3, Canada; bdeupree@ap.smu.ca

ABSTRACT

Zero age main sequence models of uniformly rotating stars have been computed for ten masses between 1.625 and 8 $M_\odot$ and 21 rotation rates from zero to nearly critical rotation. The surface shape is used to distinguish rotation rather than the surface equatorial velocity or the rotation rate. Using the surface shape is close to, but not quite equivalent to using the ratio of the rotation rate to the critical rotation rate. Using constant shape as the rotation variable means that it and the mass are separable, something that is not true for either the rotation rate or surface equatorial velocity. Thus a number of properties, including the ratio of the effective temperature anywhere on the surface to the equatorial temperature, are nearly independent of the mass of the model, as long as the rotation rate changes in such a way to keep the surface shape constant.

## 1. INTRODUCTION

Over the course of the last 60 years our knowledge of the internal structure and evolution of spherical stars has gone from rudimentary, quasi-analytic models to very detailed internal structure models which certainly explain the basic stages of stellar evolution and many, but not all, observed details. This is not to deny that a number of issues remain unresolved, but in an overall sense the state of knowledge is not bad. The same cannot be said for rotating stars. The lack of development is both theoretical and observational, but now it appears that improvements in both are bringing a better understanding of rotation in stars closer. Two observational advancements, asteroseismology and optical interferometry, are combining to provide significant constraints on rotation that can be explored with current models.

The rapid increase in observations to determine stellar oscillation frequencies from satellites such as MOST (Walker et al. 2003), CoRoT (Baglin et al. 2001), and Kepler (Basri et al. 2005) and from ground based networks such as STEPHI (Belmonte, et al. 1993) and WET (Nather et al. 1990) is generating a large number of oscillation frequencies which depend significantly on rotation if the star is rotating sufficiently rapidly. Computing oscillation modes for rotating stars has seen a significant step for rapidly rotating stars by allowing the mode to be described by more than one Legendre polynomial (Clement 1998; Lignières, Rieutord & Reese 2006; Reese, Lignières, & Rieutord 2006, 2008; Lovekin, Deupree & Clement 2009). Also, the use of models with rotation as input for these calculations (e.g., Lovekin & Deupree 2008, Reese et al. 2009) should provide more realistic oscillation frequencies for comparison with observations.

However, even with many oscillation frequencies for many stars there are simply too many degrees of freedom, at least for rapidly rotating stars, to produce rapid progress. These degrees of freedom can be significantly reduced by the development and application of optical interferometry to determine the surface shape of nearby rapidly rotating stars (e.g., van Belle et al. 2001; Domiciano de Souza et al. 2003; Monnier et al. 2007; Zhao et al. 2009). The recent application of interferometry and asteroseismology to α Oph (Monnier et al. 2010) should provide an excellent test case for learning more not only about the effects of rotation on stars but also how angular momentum is distributed in at least selected stars.

An important ingredient to comparing observations and theory is the ability to model rotating stars as realistically as possible. Two computational approaches for computing rotating stellar models were developed about 35 years ago. Both methods use the von Zeipel's theorem (1924) which shows that state variables are constant on equipotential surfaces, at least in radiative regions. One of these approaches, the self consistent field method (Ostriker & Mark 1968), solved for the gravitational potential for a given density distribution and then solved for a new density distribution on equipotential surfaces from hydrostatic equilibrium. The new densities led to a new approximation of the gravitational potential, and the entire process was iterated until all changes were sufficiently small. Jackson (1970) coupled this method to include thermal equilibrium as well and has computed a number of rapidly rotating stellar models (Jackson, MacGregor & Skumanich 2004, 2005; MacGregor et al. 2007). Bodenheimer (1971) also used the method to compute a number of rapidly rotating main sequence models for several angular momentum distributions. Clement (1974, 1978, 1979) modified the original double series expansion for the gravitational potential with a two dimensional finite difference approach.

The second approach, developed by Monaghan & Roxburgh (1965) and extended and utilized by others (e.g., Roxburgh, Griffiths & Sweet 1965; Faulker, Roxburgh & Strittmatter 1968; Kippenhahn & Thomas 1970; Sackmann & Anand 1970) allowed certain rotating models to be calculated in a one dimensional framework. This approach was extended by Endal & Sofia (1976, 1978) to include the redistribution of angular momentum in this one dimensional framework for a number of hydrodynamic and thermal instabilities, and such prescriptions are now commonly included in stellar evolution codes.

An alternative to these methods is to solve for all of the appropriate equations and their boundary conditions simultaneously on a two dimensional grid (Deupree 1990, 1995). The reason for using this framework was to include the appropriate velocity terms in the conservation equations to compute such features as meridional circulation. However, it can also be used to calculate the stellar structure of isolated models as well as members of evolution sequences.

Here I wish both to produce a large test bed of uniformly rotating models which could be used for oscillation studies and to study how the structure of models of different masses and rotation rates relate to each other. To this end I have computed a two parameter (mass and rotation) sequence of models. The first issue is to decide what the rotation variable will be, and I selected the surface shape for this. The rationale is that models with the same surface shape have the same ratio of centrifugal potential to gravitational potential, so that the effect of rotation, at least at the surface, would be the same. Therefore, I can examine models of different masses but with the same surface shape and models with the same mass but different surface shapes. It turns out that the surface shape was a significant choice as the rotation variable because mass and the surface shape are essentially separable variables, something that is not true for either mass and rotation rate or mass and surface equatorial velocity. I will show that the same surface shape is close to, but not exactly the same as constant ratio between the rotation rate and the critical rotation rate. I discuss a few details about the computation of the rotating models in Section 2, and present the results for the two dimensional array of rotating models in Section 3. Section 4 is a brief discussion of the results.

## 2. COMPUTATION OF ROTATING MODELS

I wish to summarize briefly the major features of the computation of the 2D rotating ZAMS models. A model is computed on a 2D finite difference grid whose independent variables are the fractional surface equatorial radius (x) and the colatitude ($\theta$). A rotation rate is imposed as a function of x and $\theta$ in terms of the azimuthal velocity, $v_\varphi$. In this paper the rotation is taken as uniform so the azimuthal velocity increases linearly with the distance from the rotation axis and the magnitude is determined by the surface equatorial azimuthal velocity. The composition is taken to be X = 0.7, Z = 0.02 and assumed to be uniform throughout the ZAMS model. The opacity and equation of state are computed from the OPAL tables (Rogers, Swenson & Iglesias 1996; Iglesias & Rogers 1996) and the composite energy generation rates for both p-p and CNO hydrogen burning are taken from Fowler, Caughlan & Zimmermann (1967). Time independent equations of hydrostatic and thermal equilibrium, along with Poisson's equation, the equation of state and relations for the nuclear energy generation and the radiative opacity are solved simultaneously for the density, pressure, temperature and gravitational potential at each location in the 2D grid. The thermal balance equation assumes radiative diffusion, with the opacity modified as discussed below to produce an adiabatic gradient in convective cores. The method of solution is a two dimensionalization of the Henyey technique outlined by Deupree (1990). The determination of the surface equatorial radius is found by integrating the density over the volume to produce the imposed mass, an equation which is also solved as part of the Henyey method. Azimuthal and equatorial symmetry are imposed to supply the horizontal boundary conditions.

The surface provides some interesting challenges. First, a procedure must be found to determine its location at each angle ($x_{surf}(\theta)$). Once the surface is found we impose the boundary conditions of the surface density being some very small value and the surface temperature being equal to the effective temperature divided by the fourth root of two. This latter relation comes from applying the simple Eddington atmosphere

$$T^4(\tau) = \frac{3}{4} T_{eff}^4 \left( \tau + \frac{2}{3} \right)$$

at $\tau = 0$. Such a relationship between the effective temperature and the local temperature is contradictory to von Zeipel's (1924) law which forces the temperature to be constant on an equipotential surface (including the stellar surface), even though the effective temperature can vary by a significant factor between the pole and equator. This effect is small at low rotation because the equatorial to polar effective temperature variation is small, but the differences between von Zeipel's law and the surface temperature structure do become noticeable once the ratio of the polar to equatorial effective temperature becomes about 10%. It does highlight some of the uncertainty associated with the surfaces of rotating models. Interestingly, this alters the ratio of the polar to equatorial radiative flux by about the same amount as the hydrodynamic simulations of Espinosa Lara & Rieutord (2007) in comparison to the flux ratio of models with von Zeipel's law applied all the way to the surface.

The convective core is assumed to be adiabatic. Because of my desire to compute hydrodynamic simulations with this code, I imposed this condition by replacing the opacity in the radiative flux in the energy conservation equation by an effective opacity

$$\kappa_{eff} = \left(\frac{\nabla_{ad}}{\nabla_{rad}}\right)\kappa$$

in the convective core. Convectively unstable regions associated with ionization zones are quite inefficient and small in the models considered here, and convection in them has been ignored. It is true that the equatorial regions of rapidly rotating models reach significantly lower effective temperatures than the nonrotating models, but the conditions in the hydrogen ionization region there are quite similar to those in the nonrotating model. This happens because the cooler effective temperatures are offset by the significantly lower gravity. There is also an issue with the radiative gradient that needs to be considered. Recall that the radiative gradient is computed by dividing the temperature gradient produced by radiative transport by the pressure gradient produced by hydrostatic equilibrium. Of course, the usual expression for hydrostatic equilibrium does not include rotation, so I compute the radiative gradient by dividing the expression for the radial component of the temperature gradient by the finite difference expression for the radial pressure gradient. This different radiative gradient has the capability to change the location of the convective core boundary (determined by the equality of this expression for the radiative gradient and the adiabatic gradient), but in fact the change is negligible unless the rotation rate near the convective core boundary is large. This only happens if the rotation rate increases significantly as the distance from the rotation axis decreases, something not true in the calculations discussed in this work.

Because I use a finite difference non-Lagrangian framework, the locations of both the convective core boundary and the surface are quantized – for example, a computational cell is either a surface zone or it is not, so that the surface will be at the same place for a limited set of conditions until the change in those conditions becomes sufficiently large to move the equipotential (for the surface) or the equality of the radiative and adiabatic gradients (for the convective core boundary) into one of the two neighboring radial cells.

## 3. MODELS WITH ROTATION

It should be easier to relate the effects of rotation on models at different masses if the effect of rotation is the same for all masses. As mentioned previously, I felt the best way to do this in a general sense is to use the constant surface shape for all masses to be the rotation parameter. This is clearly better than using either the rotation rate or the surface equatorial velocity as the rotation parameter because the effect rotation has on the model changes as a function of mass for these two. If the gravitational potential was given by the Roche potential, then constant surface shape would be closely related to constant S, defined by

$$S^2 = \frac{\Omega^2 R_{eq}^3}{GM}$$

The parameter S is related to the ratio of the rotation rate ($\Omega$) to the critical rotation rate ($\Omega_{crit}$), which is that rotation rate for which the centrifugal force will balance gravity (i.e., S = 1). However, neither the critical rotation rate nor the surface equatorial radius ($R_{eq}$) at critical rotation is known a priori, whereas S can be computed for any given model available. One of the features I wish to examine is how good a proxy S is for constant surface shape.

I have computed 21 rotating models (including zero rotation) for ten different masses ranging from 1.625 to 8 $M_\odot$. The shapes used to control the rotation variation correspond to surface equatorial velocities for the uniformly rotating 10 $M_\odot$ models used by Lovekin, Deupree, & Clement (2009) and Deupree & Beslin (2010) of 0, 10, 30, 50, 90, 120, 150, 180, 210, 240, 270, 300, 330, 360, 390, 420, 450, 480, 510, 540, and 570 km s$^{-1}$. All models had 581 radial zones and ten angular zones. Convergence at lower masses and higher rotation rates was hindered in some cases by the very steep temperature gradient in the hydrogen ionization region and possibly by the radial numerical resolution in this region at certain angles. The problem is that changing the surface location by a zone requires mapping of the dependent variables between some arbitrarily chosen depth inside the model and the new surface. Because many quantities are varying rapidly with radius near the surface, particularly if there is a hydrogen ionization region with inefficient convection, obtaining a good approximation to the new structure can be difficult. The failure to obtain a sufficiently accurate first guess to the new structure can lead to convergence difficulties. This problem was eliminated by a capability I built into ROTORC which allows the user to set all perturbations equal to zero inside some user selected radial zone number. Once it became clear that convergence using the full model was not happening, I began to converge the model from the hydrogen ionization zone outward and then gradually working in to deeper layers until all zones in the model were included. In all models both the density and temperature are converged to about one part in $10^8$ (this converges the gravitational potential to approximately the round off error of a few parts in $10^{15}$).

I present a summary of the basic properties of the ZAMS nonrotating models in Table 1, along with some rotation information. The first columns are the mass, the radius, effective temperature, convective core mass, central density, and central temperature of the nonrotating models. As I will discuss shortly, the models show that ratios of the rotation rate, the surface equatorial velocity, and the equatorial effective temperature between two masses as nearly independent of the rotation rate. These three quantities are presented as fractions of the respective quantities for the 8 $M_\odot$ model in the last three columns. Before examining details of the models, I will explore the relationship among three possible representations of the rotation parameter: surface shape, S, and $\Omega / \Omega_{crit}$.

*3.1 Relation between Possible Rotation Parameters*

I first examine the relationship between S and models with constant surface shape. For each surface shape, I compute S for each mass and take the average over all the masses. These values of S, along with the similarly computed values of $\Omega / \Omega_{crit}$, are presented in Table 2 for each of the constant surface shapes used. These shapes are denoted by "r" in the table with an integer between 1 and 20 (0 corresponds to no rotation and is omitted). The variation of S is typically about 0.0005, almost independent of rotation, while the variation of $\Omega / \Omega_{crit}$ is about 0.002. The trend in all cases is for both S and $\Omega / \Omega_{crit}$ to increase as the mass decreases, although it may not always be true from one mass to the next. I show the fractional variation of S with respect to the 8 $M_\odot$ value as a function of mass for several surface shapes in Figure 1. Generally speaking, the results show that S is not a bad proxy for constant surface shape, but it is not exact. However, S does have the distinct advantage in that it provides a quantitative determination of the effects of rotation. Furthermore, preliminary calculations show that S is not nearly as greatly affected by differential rotation as is the simplest measurement of constant shape, the ratio of the surface polar radius to the surface equatorial radius (see Figures 2 and 3 of Lovekin, Deupree & Clement (2009) for the sensitivity of this ratio to differential rotation). The behavior of S with differential rotation is beyond the scope of the current work. I will use S (or $S^2$) in plots because it conveys quantitative information, although the surface shape is really the parameter I am using. The behavior of $\Omega / \Omega_{crit}$ with mass will be discussed shortly.

This variation of S is presumably related to the differences in the nonspherical variation of the gravitational potential at different masses. In Figure 2 I present the difference between the gravitational potential along the polar axis and the gravitational potential along the equator divided by the equatorial gravitational potential. This is shown as a function of the fractional surface equatorial radius for the 1.625 $M_\odot$ and the 8 $M_\odot$ models rotating close to critical rotation. The maximum amplitude of this difference is larger for the higher mass model by almost a factor of 2.5, although the total variation in both cases remains less than one per cent. The curves for more slowly rotating models have the same shape, but with lower amplitudes, and the ratio of the maximum amplitudes between the high and low mass model is higher for lower rotation. Figure 2 also indicates where the major perturbations to the mass occur. Close to the center of the model the rotation is too small to significantly alter the mass distribution. There is little mass exterior to the minima and the decreased effect of the variations with increasing fractional radius occurs because the significantly perturbed mass is farther away. For example, only 2.5% of the mass is beyond a fractional surface equatorial radius of 0.4 for the 8 M$_\odot$ model. The fractional surface equatorial radius beyond which there is only 2.5% of the mass is about 0.33 for the 1.625 M$_\odot$ model. The equivalent locations in the nonrotating models are 0.55 for the 8 M$_\odot$ model and 0.475 for the 1.625 M$_\odot$ model. The smaller degree of concentration for the higher mass star is consistent with the fact that the convective core with its smaller effective polytropic index and hence lesser mass concentration contains about 29% of the mass of the more massive model and only about 10% for the less massive model.

The ratio of the rotation rate to the critical rotation rate is often used as an indicator of the importance of rotation. I show this as a function of S in Figure 3 for the 1.625 M$_\odot$ and the 8 M$_\odot$ model rotation sequences, where it is clear that S and $\Omega / \Omega_{crit}$ are close, but not identical for the two masses. Neither is exactly the same as constant surface shape, but all three are close to each other, at least for these uniform rotation models. One interesting feature about Figure 3 is the relationship between S and $\Omega / \Omega_{crit}$, in that high values of $\Omega / \Omega_{crit}$ are reached at quite moderate values of S. To me this makes S preferable to $\Omega / \Omega_{crit}$ for representing the effects of rotation.

We can understand the mass dependence of S seen in Fig. 1 in terms of the departures ($\delta$ and $\varepsilon$) of the actual gravitational potential from the Roche potential at the pole and equator, respectively. We start with the equipotential ($\Psi$) at the model surface for the pole and the equator

$$\Psi = \Phi_{eq} - \frac{V_{eq}^2}{2} = (1+\varepsilon)\frac{GM}{R_{eq}} - \frac{V_{eq}^2}{2} = \Phi_{pole} = (1+\delta)\frac{GM}{R_{pole}}$$

M is the model mass, $V_{eq}$ is the surface equatorial velocity, and $R_{eq}$ and $R_{pole}$ are the surface radii at the equator and pole, respectively. Even for the most oblate model, the amount of mass outside the polar radius is less than one part in $10^5$ and the effect is small compared to that of $\delta$ and $\varepsilon$. Dividing by GM / R and using the definition of S leads to

$$\frac{S^2}{2} = (1+\delta)\frac{R_{eq}}{R_{pole}} - (1+\varepsilon)$$

Because of the elongation of the model in the equatorial direction, $\varepsilon$ is slightly positive, while $\delta$ is larger in magnitude and negative. The difference in magnitude is due to the fact that the equator is farther away from the mass concentration (once the star is rotating fast enough); $\varepsilon$ is positive because the mass which is disturbed by rotation has been moved outward on the equatorial plane, and $\delta$ negative because the disturbed mass has been moved farther away from the polar axis. From Fig. 2 the magnitudes of $\delta$ and $\varepsilon$ are larger for higher mass models. In the equation above, the effect of $\delta$ is larger, leaving a smaller value of S for larger magnitudes of $\delta$.

## 3.2 Variation of Properties with Mass

The last three columns in Table 1 show the rotation rate, the surface equatorial velocity, and the equatorial effective temperature as functions of mass. All three quantities have been scaled to their values for the 8 $M_\odot$ model. The variation of these ratios with S is small. The standard deviation from the average over all nonzero r values for the rotation rate ratio is between 0.01 and 0.017 and increases as the mass decreases, as one might expect because of the scaling to the 8 $M_\odot$ model. The standard deviation for the surface equatorial velocity is about 0.0035 and does not vary appreciably from one mass to the next. The standard deviation for the equatorial effective temperature is only about 0.0003 and also shows no significant dependence on mass. Figure 4 presents the variation of the rotation rate with mass for four r values.

An interesting quantity which is almost independent of mass is the ratio between the polar and equatorial effective temperatures for a given value of r. I show a fairly typical example in Figure 5, where it is seen that the ratio tends to decrease with mass, but that the decrease is comparatively small. It turns out that the small variation in effective temperature ratio is true not only for the ratio of the polar to equatorial effective temperature, but also for the ratio between the effective temperature at any latitude and the equatorial effective temperature. This has the significant result that some modest knowledge about the rotation can lead to a reasonable deduction of the surface effective temperature as a function of latitude, at least for uniform rotation. It is reasonable to expect, although not proven, that this near independence of the effective temperature ratios at different latitudes will be maintained with other conservative rotation laws, although the precise values for the ratios will depend on the rotation law.

## 3.3 Variation of Properties with Rotation

I have indicated that a number of properties are almost independent of mass, so it remains to determine how they vary with rotation. The simplest of these is the surface shape, which is independent of mass by definition. I have also shown that the ratio of the polar to equatorial effective temperature is nearly independent of mass and indicated that this is true at other latitudes besides the pole. In Table 3 I present the surface radius and effective temperature, in units of their equatorial values, as functions of colatitude for the values of the rotation parameter r. I have omitted r0 for obvious reasons and r1 because the variations are too small for the number of significant digits in the table. The near independence of the effective temperature as a function of mass can be understood from von Zeipel's (1924) law. The radiative flux is perpendicular to the equipotential surfaces, whose shape (at least at the surface) is independent of mass, so it is not unreasonable to suspect that the common surface shape leads to the same flux distribution with latitude. Even the violation of von Zeipel's law near the surface may not matter either because it occurs only close to the surface or because it is violated in the same way for all masses. It should be noted that the near independence of the effective temperature ratio with mass is due to a given rotation being defined in terms of a given surface shape. If the shape changes at any latitude, the effective temperature must change to deliver the flux that is coming from the interior.

In order to complete the description of the properties in Tables 1 and 3 for all masses and rotation rates, I must indicate their variation with r. This is done in Table 4 and includes the rotation rate and the surface equatorial velocity for the 8 M$_\odot$ model, the surface equatorial radius, and the equatorial effective temperature. The values given for the last two quantities, presented as ratios to their nonrotating values, are the average over all the masses at a given rotation. The variation in the surface equatorial radius is shown in Figure 6 for the 1.625 and 8 M$_\odot$ models. There is some variation in the relationship for the different masses, particularly noticeable for higher rotation. The variation between the two curves is smooth as a function of mass.

This variation explains the variation of $\Omega / \Omega_{crit}$ with mass. From Figure 6 we can write

$$\frac{\delta R_{eq}}{R_{eq}} = aS^2 = a\frac{\Omega^2 R_{eq}^3}{GM} = a\frac{\Omega^2 R^3 (1+aS^2)}{GM}$$

Here R is the surface radius of the nonrotating model. Using the fact that $S^2 = 1$ for critical rotation, I can divide this equation by the similar expression for critical rotation and solve for the ratio of the rotation rates to obtain

$$\left(\frac{\Omega}{\Omega_{crit}}\right)^2 = \frac{(1+a^3)S^2}{(1+aS^2)^3}$$

The right hand side is larger for larger a, which corresponds to lower masses in Figure 6.

A similar plot is shown for the equatorial effective temperature in Figure 7. There is one interesting difference between this plot and Figure 6; the variation between the two curves is smooth except for the two most rapidly rotating data points. Here the 1.75 M$_\odot$ data is about halfway between the 1.625 and 8 M$_\odot$ results (effectively a continuation of the relationship at slower rotation), and the other masses fall in between the 1.75 and 8 M$_\odot$ results. It is not clear why these high rotation results for the 1.625 M$_\odot$ model have this behavior.

The surface polar quantities are of interest because it is often assumed that the polar radius remains unchanged from its nonrotating value (e.g., Collins, Truax & Cranmer 1991; Porter 1996; Reiners 2003; Townsend, Owocki & Howarth 2004), although this assumption is not universal (e.g., Frémat, et al. 2005). Indeed, the fractional changes in both the polar effective temperature and polar radius are comparatively small, as is shown in Figures 8 and 9, respectively. In Figure 8 the point for the most rapidly rotating 1.625 M$_\odot$ model is again different from that of other masses, which show the same slope between the two most rapidly rotating models as does the 8 M$_\odot$ model. The magnitude of the polar radius changes shown in Figure 9 are a little higher than the (Roche potential) results of Sackmann (1970) and those from calculations in the relevant mass range (Faulkner, Roxburgh & Strittmatter 1968; Sackmann & Anand 1969;

Kippenhahn & Thomas 1970) to which she compared. Given the previous discussion, this can be attributed to my computation of the actual potential. The trend with mass is also the same with all these calculations except those of Sackmann and Anand (1969), which for some reason show an increase with decreasing mass. The primary differences between the calculations of Sackmann (1970) and Sackmann and Anand (1969) relate to the opacity used and the surface boundary conditions.

### 3.4 Variation of Internal Properties with Rotation

It is well known that the central properties of uniformly rotating models are only slightly affected by rotation because critical rotation is reached before the core rotation rates are sufficiently large for the centrifugal force to become even a modest fraction of the gravitational force. The decrease in central temperature at critical rotation is only 0.4% for the 8 $M_\odot$ model and 0.6% for the 1.625 $M_\odot$ model, with the other masses being in between. The change is approximately linear through model r16, with very little change for more rapidly rotating models. The central density increases somewhat more, as is shown in Figure 10 as the ratio of the central density to that in the nonrotating model as a function of S. I have shown the results for all masses here to give an impression of the behavior with mass, which could not be done in most of other plots. The mass of the convective core is unchanged within the quantization produced by the zoning discussed in the previous section. The flux emerging from the convective core is essentially spherically symmetric, and the luminosity flowing through a spherical shell just outside the core is shown in Figure 11 as a function of S. These results are in close agreement with those of other authors in the comparison performed by Sackmann (1970). These trends in the central density, central temperature, and luminosity are generally explained (e.g., Sackmann 1970) by the primary effect of rotation on the model structure being to reduce the effective gravity, so that an individual rotating model would have the properties of a nonrotating model with a slightly smaller mass. Although the change in the luminosity produced by the core is small, the luminosity an observer would deduce assuming spherical symmetry would vary appreciably as a function of the observer's inclination to the rotation axis (e.g., Collins 1966; Maeder & Peytremann 1970; Slettebak, Kuzma, & Collins 1980; Linnell & Hubeny 1994; Reiners & Schmitt 2002; Gillich, et al. 2008). The same is true of the effective temperature, although one can define an average effective temperature using the luminosity and the total surface area. This average effective temperature is about 1900 K cooler for a 6 $M_\odot$ model rotating near critical rotation than for the nonrotating model, a value not far removed from the 1D models of Brott, et al. (2011).

I have looked at a number of ways to compare the density and temperature distributions of rotating and nonrotating models. The comparison which produces the best results is comparing the nonrotating model density and temperature to the density and temperature along the polar axis as functions of the fractional polar surface radius. I show the results of this comparison in Figure 12 as the ratio of the temperature on the polar axis for model r20 to that of the nonrotating model and the same for the density. The differences for both variables are less than 3% out to a fractional polar radius of about 0.75, after which they rise significantly towards the surface. There does not appear to be any easy method of relating the properties near the surface because the temperature and density vary so rapidly as a function of radius. The curves for other masses are basically the same shape, although the amplitudes are reduced for lower masses.

## 4. DISCUSSION

I have computed a number of ZAMS 2D stellar models with solid body rotation covering a range of mass and rotation rates. I have found that the mass and rotation are separable variables if the rotation variable is defined in such a way that the variation from one mass to another keeps the surface shape constant. The rotation variable, S, required to do this for uniformly rotating models is close to, but nor identical to, the ratio of the centrifugal potential to the Roche potential. The failure to be identical is because the gravitational potential deviates slightly from the Roche potential. A number of quantities, including the surface equatorial velocity, the rotation rate, the surface equatorial velocity, and the equatorial effective temperature, scale from one mass to another (as long as the surface shape is not changed while the mass changes) by factors that are themselves nearly independent of the surface shape. The latitudinal variation of the surface radius (by design) and of the effective temperature in units of their equatorial values depend on the surface shape, but are nearly independent of mass for a given surface shape. This independence will allow the calculation of the apparent spectral and photometric properties one would observe for uniformly rotating models without having to compute the rotating models throughout the HR diagram.

While not proven, it is reasonable to expect this independence will remain true for other conservative rotation laws. However, it is likely that the differences between constant surface, S, and $\Omega / \Omega_c$ will become larger for differential rotation in which the rotation rate increases as the distance from the rotation axis decreases because the magnitude of the deviation of the gravitational potential from the Roche potential will become larger.

All these models are on the ZAMS, and the conclusions only apply strictly to this stage. As the star ages thermal and hydrodynamic instabilities associated with rotation will cause the models to mix both composition and angular momentum, and variations in these from one model to another will have to be considered. The results presented here might be of benefit in comparing the effects of properties such as the total mass or internal composition distribution while minimizing the effects of the rotational structure. These results would indicate that comparing models with the same surface shape, assuming that the rotation laws are conservative, is a reasonable way to do this.

I wish to thank the anonymous referee for his questions which led to the analysis about the effects of the non-Roche potential.

# TABLE 1

## MODEL PROPERTIES AS FUNCTIONS OF MASS

| $M/M_\odot$ | $R/R_\odot$ | $T_{eff}$ (K) | $M_{cc}/M_\odot$ | $\rho_c$ g cm$^{-3}$ | $T_c / 10^7$ (K) | $\Omega/\Omega_8$ | $V_{eq}/V_{eq,8}$ | $T_{eff}/T_{eff,8}$ |
|---|---|---|---|---|---|---|---|---|
| 8……    | 3.505 | 22,430 | 2.332 | 11.435 | 3.008 | 1.0000 | 1.0000 | 1.0000 |
| 6……    | 2.976 | 19,029 | 1.613 | 16.090 | 2.829 | 1.1049 | 0.9401 | 0.8486 |
| 4……    | 2.363 | 14,814 | 0.901 | 26.803 | 2.577 | 1.2748 | 0.8618 | 0.6608 |
| 3……    | 2.007 | 12,252 | 0.606 | 38.951 | 2.400 | 1.4099 | 0.8105 | 0.5465 |
| 2.5….   | 1.812 | 10,810 | 0.466 | 49.245 | 2.285 | 1.4988 | 0.7786 | 0.4821 |
| 2.25…  | 1.711 | 10,034 | 0.389 | 56.191 | 2.216 | 1.5649 | 0.7600 | 0.4476 |
| 2……    | 1.610 | 9,206  | 0.309 | 64.667 | 2.135 | 1.5991 | 0.7386 | 0.4107 |
| 1.875.. | 1.560 | 8,766  | 0.263 | 69.528 | 2.087 | 1.6204 | 0.7263 | 0.3916 |
| 1.75…  | 1.513 | 8,301  | 0.213 | 74.744 | 2.032 | 1.6394 | 0.7125 | 0.3707 |
| 1.625... | 1.470 | 7,805  | 0.165 | 80.121 | 1.966 | 1.6496 | 0.6967 | 0.3489 |

TABLE 2

MEASURES OF ROTATION

| r | $\langle S \rangle$ | $\langle \Omega / \Omega_{crit} \rangle$ |
|---|---|---|
| 1 | 0.01165 | 0.0204 |
| 2 | 0.04353 | 0.0772 |
| 3 | 0.07314 | 0.1295 |
| 4 | 0.13163 | 0.2311 |
| 5 | 0.17484 | 0.3044 |
| 6 | 0.22001 | 0.3785 |
| 7 | 0.26601 | 0.4514 |
| 8 | 0.31114 | 0.5189 |
| 9 | 0.35649 | 0.5829 |
| 10 | 0.40486 | 0.6470 |
| 11 | 0.45388 | 0.7065 |
| 12 | 0.50271 | 0.7600 |
| 13 | 0.55464 | 0.8104 |
| 14 | 0.60883 | 0.8564 |
| 15 | 0.66300 | 0.8951 |
| 16 | 0.72134 | 0.9294 |
| 17 | 0.78222 | 0.9571 |
| 18 | 0.84726 | 0.9792 |
| 19 | 0.91594 | 0.9934 |
| 20 | 0.98991 | 1.0000 |

TABLE 3

VARIATION OF $T_{eff}$ AND R AT DIFFERENT COLATITUDES AS FUNCTIONS OF ROTATION ID

| r ID | 4.5 | 13.5 | 22.5 | 31.5 | 40.5 | 49.5 | 58.5 | 67.5 | 76.5 |
|---|---|---|---|---|---|---|---|---|---|
| | | | | $T_{eff}(\theta) / T_{eff}(\theta = 85.5)$ | | | | | |
| r2 | 1.0010 | 1.0010 | 1.0009 | 1.0007 | 1.0005 | 1.0004 | 1.0003 | 1.0001 | 1.0000 |
| r3 | 1.0027 | 1.0026 | 1.0024 | 1.0021 | 1.0016 | 1.0012 | 1.0008 | 1.0004 | 1.0002 |
| r4 | 1.0089 | 1.0088 | 1.0076 | 1.0065 | 1.0053 | 1.0038 | 1.0025 | 1.0013 | 1.0005 |
| r5 | 1.0157 | 1.0149 | 1.0136 | 1.0117 | 1.0093 | 1.0069 | 1.0044 | 1.0024 | 1.0008 |
| r6 | 1.0247 | 1.0233 | 1.0216 | 1.0181 | 1.0147 | 1.0110 | 1.0073 | 1.0038 | 1.0013 |
| r7 | 1.0348 | 1.0339 | 1.0310 | 1.0262 | 1.0218 | 1.0165 | 1.0107 | 1.0058 | 1.0022 |
| r8 | 1.0475 | 1.0465 | 1.0428 | 1.0351 | 1.0298 | 1.0226 | 1.0149 | 1.0084 | 1.0030 |
| r9 | 1.0638 | 1.0597 | 1.0549 | 1.0473 | 1.0387 | 1.0295 | 1.0205 | 1.0113 | 1.0041 |
| r10 | 1.0796 | 1.0775 | 1.0696 | 1.0624 | 1.0513 | 1.0381 | 1.0268 | 1.0151 | 1.0056 |
| r11 | 1.0992 | 1.0956 | 1.0906 | 1.0763 | 1.0656 | 1.0485 | 1.0334 | 1.0202 | 1.0075 |
| r12 | 1.1226 | 1.1190 | 1.1057 | 1.0982 | 1.0785 | 1.0630 | 1.0419 | 1.0265 | 1.0099 |
| r13 | 1.1480 | 1.1420 | 1.1327 | 1.1152 | 1.1003 | 1.0783 | 1.0538 | 1.0333 | 1.0136 |
| r14 | 1.1764 | 1.1687 | 1.1558 | 1.1406 | 1.1203 | 1.0961 | 1.0701 | 1.0430 | 1.0186 |
| r15 | 1.2047 | 1.1931 | 1.1860 | 1.1645 | 1.1444 | 1.1178 | 1.0862 | 1.0511 | 1.0244 |
| r16 | 1.2429 | 1.2283 | 1.2170 | 1.1979 | 1.1788 | 1.1424 | 1.1112 | 1.0682 | 1.0340 |
| r17 | 1.2736 | 1.2721 | 1.2558 | 1.2296 | 1.2117 | 1.1791 | 1.1353 | 1.0889 | 1.0462 |
| r18 | 1.3188 | 1.3160 | 1.2932 | 1.2735 | 1.2464 | 1.2105 | 1.1710 | 1.1172 | 1.0644 |
| r19 | 1.3585 | 1.3498 | 1.3361 | 1.3236 | 1.2817 | 1.2546 | 1.2044 | 1.1512 | 1.0841 |
| r20 | 1.3965 | 1.4065 | 1.3866 | 1.3657 | 1.3261 | 1.2964 | 1.2434 | 1.1916 | 1.1160 |
| | | | | $R(\theta) / R(\theta = 85.5)$ | | | | | |
| r2 | 0.9990 | 0.9991 | 0.9992 | 0.9993 | 0.9995 | 0.9996 | 0.9997 | 0.9998 | 0.9999 |
| r3 | 0.9974 | 0.9975 | 0.9977 | 0.9980 | 0.9985 | 0.9989 | 0.9993 | 0.9996 | 0.9999 |
| r4 | 0.9915 | 0.9918 | 0.9927 | 0.9938 | 0.9950 | 0.9964 | 0.9977 | 0.9988 | 0.9996 |
| r5 | 0.9850 | 0.9858 | 0.9871 | 0.9889 | 0.9911 | 0.9935 | 0.9959 | 0.9978 | 0.9992 |
| r6 | 0.9764 | 0.9776 | 0.9795 | 0.9826 | 0.9861 | 0.9897 | 0.9933 | 0.9964 | 0.9988 |
| r7 | 0.9662 | 0.9674 | 0.9701 | 0.9746 | 0.9795 | 0.9847 | 0.9901 | 0.9947 | 0.9981 |
| r8 | 0.9542 | 0.9558 | 0.9596 | 0.9656 | 0.9717 | 0.9792 | 0.9863 | 0.9926 | 0.9974 |
| r9 | 0.9403 | 0.9434 | 0.9482 | 0.9550 | 0.9630 | 0.9722 | 0.9817 | 0.9901 | 0.9965 |
| r10 | 0.9250 | 0.9276 | 0.9337 | 0.9414 | 0.9517 | 0.9637 | 0.9759 | 0.9868 | 0.9953 |
| r11 | 0.9074 | 0.9106 | 0.9167 | 0.9276 | 0.9393 | 0.9542 | 0.9691 | 0.9830 | 0.9939 |
| r12 | 0.8878 | 0.8916 | 0.9007 | 0.9106 | 0.9263 | 0.9424 | 0.9610 | 0.9780 | 0.9921 |
| r13 | 0.8663 | 0.8709 | 0.8797 | 0.8935 | 0.9091 | 0.9289 | 0.9509 | 0.9722 | 0.9897 |
| r14 | 0.8436 | 0.8490 | 0.8592 | 0.8732 | 0.8916 | 0.9137 | 0.9382 | 0.9643 | 0.9866 |
| r15 | 0.8202 | 0.8263 | 0.8352 | 0.516 | 0.8709 | 0.8954 | 0.9237 | 0.9550 | 0.9823 |
| r16 | 0.7933 | 0.8003 | 0.8105 | 0.8263 | 0.8463 | 0.8754 | 0.9058 | 0.9424 | 0.9768 |
| r17 | 0.7665 | 0.7705 | 0.7822 | 0.8003 | 0.8202 | 0.8490 | 0.8858 | 0.9263 | 0.9685 |
| r18 | 0.7362 | 0.7407 | 0.7539 | 0.7705 | 0.7933 | 0.8233 | 0.8592 | 0.9041 | 0.9550 |
| r19 | 0.7071 | 0.7122 | 0.7220 | 0.7362 | 0.7623 | 0.7896 | 0.8293 | 0.8754 | 0.9337 |
| r20 | 0.6749 | 0.6749 | 0.6860 | 0.7020 | 0.7268 | 0.7539 | 0.7933 | 0.8381 | 0.8990 |

TABLE 4

MODEL PROPERTIES AS FUNCTIONS OF ROTATION ID

| r ID | $\Omega_8$ ($10^{-6}$ rad s$^{-1}$) | $V_{eq,8}$ (km s$^{-1}$) | $R_{eq}$ (rID) / $R_{eq}$ (r0) | $T_{eq}$ (rID) / $T_{eq}$ (r0) |
|---|---|---|---|---|
| r1 | 3.689 | 9. | 1.00008 | 0.99994 |
| r2 | 11.714 | 28.6 | 1.00102 | 0.99933 |
| r3 | 19.632 | 48. | 1.00246 | 0.99812 |
| r4 | 35.004 | 86. | 1.00767 | 0.99394 |
| r5 | 46.152 | 114. | 1.01355 | 0.98935 |
| r6 | 57.471 | 143. | 1.02134 | 0.98327 |
| r7 | 68.115 | 171. | 1.03114 | 0.97579 |
| r8 | 78.911 | 200.3 | 1.04262 | 0.96726 |
| r9 | 89.094 | 229. | 1.05596 | 0.95764 |
| r10 | 98.815 | 257.7 | 1.07251 | 0.94578 |
| r11 | 107.966 | 286.3 | 1.09143 | 0.93280 |
| r12 | 116.290 | 314.2 | 1.11282 | 0.91852 |
| r13 | 124.229 | 343. | 1.13833 | 0.90212 |
| r14 | 131.375 | 371.6 | 1.16830 | 0.88615 |
| r15 | 137.692 | 400.2 | 1.19962 | 0.86491 |
| r16 | 143.111 | 428.8 | 1.23751 | 0.84335 |
| r17 | 147.420 | 457. | 1.28089 | 0.81992 |
| r18 | 150.951 | 485.7 | 1.33048 | 0.79485 |
| r19 | 153.275 | 514.3 | 1.38793 | 0.77014 |
| r20 | 154.394 | 543.2 | 1.45520 | 0.74490 |

Figure Captions

Fig. 1 – The ratio of S to its value for the 8 $M_\odot$ model as a function of mass in $M_\odot$ for rotation sequences r4 (x's), r8 (diamonds), r12 (squares), and r19 (circles). S generally increases as the mass decreases. Neither this ratio nor the absolute variation of S is independent of rotation sequence.

Fig. 2 – Fractional difference between the gravitational potential along the polar axis and that along the equatorial plane as a function of the fractional surface equatorial radius for models near critical rotation. The solid curve is for 1.625 $M_\odot$, and the dashed curve is for 8 $M_\odot$. The fractional difference is larger for larger masses and the maximum magnitude is farther out in terms of fractional equatorial radius.

Fig. 3 – Ratio of the rotation rate to the critical rotation rate as a function of S for an 8 $M_\odot$ rotation sequence (solid curve) and for a 1.625 $M_\odot$ rotation sequence (dashed curve). The two curves are close to each other, but not identical. Neither S nor $\Omega / \Omega_{crit}$ are identical to constant surface shape, but all three are close to each other, at least for uniform rotation.

Fig. 4 - Rotation rate, scaled to the rotation rate for the 8 $M_\odot$ model, versus mass in $M_\odot$ for four rotation sequences: r5 (upper curve), r10, r15, and r20 (lower curve, near critical rotation). For a given rotation as identified by constant surface shape, the rotation rate increases as the mass decreases, but the ratio of the rotation rate between one mass and another is not strongly dependent of the amount of rotation, at least for the mass range covered here.

Fig. 5 – Ratio of the effective temperature at the pole to the effective temperature at the equator as a function of mass in $M_\odot$ for rotation r14. Both the tendency for the ratio to increase with increasing mass and the small total variation are typical. Of course, the value of the ratio increases with increasing rotation.

Fig. 6 – Increase in the surface equatorial radius in units of the surface radius of the nonrotating model as a function of $S^2$ for 8 $M_\odot$ (circles) and 1.625 $M_\odot$ (squares). The solid line and dashed line are linear least squares fits to the two data sets and fit the data well. Note that there is some dependence on mass in this relationship.

Fig. 7 – Decrease in the equatorial effective temperature as a fraction of the effective temperature of the nonrotating model as a function of $S^2$ for 8 $M_\odot$ (circles) and 1.625 $M_\odot$ (squares). The solid and dashed lines are merely drawn between individual points. There is some dependence on mass in this relationship.

Fig. 8 – Variation in the effective temperature at the pole in units of the effective temperature of the nonrotating model as a function of $S^2$. The circles are data for 8 $M_\odot$ and the squares are data for 1.625 $M_\odot$. There is some variation with mass, but a good approximation is to assume a linear increase with $S^2$ up to about 0.4 and constant at about 0.045 for larger values of $S^2$.

Fig. 9 – Variation in the polar surface radius in units of the surface radius of the nonrotating model. Circles represent data for 8 $M_\odot$ and squares for 1.625 $M_\odot$. This is one of the few quantities for which there is a noticeable difference as a function of mass.

Fig. 10 – Variation in the central density in units of the central density of the nonrotating model as a function of S. All ten masses from Table 1 are included, with the percentage increase being smallest for the 1.625 $M_\odot$ models (*) and largest for the 8 $M_\odot$ models (o). The increase is comparatively small for all models, reflecting the truism that models with uniform rotation reach critical rotation before the rotation rate is sufficiently large to affect the core appreciably.

Fig. 11 – The change in luminosity emerging from the core in units of the luminosity of the nonrotating model for all ten masses in Table 1 as a function of S. The 8 $M_\odot$ models (o) are affected most and the 1.625 $M_\odot$ models least (*). Because of the latitudinal dependence of the flux emerging from the surface, an observer assuming a nonrotating model might deduce a luminosity quite different from that actually being produced.

Fig. 12 – The solid line is the ratio of the temperature at a given fractional polar radius for a model rotating near critical rotation (r20) to the temperature at the same fractional radius for a nonrotating model (r0) as a function of the fractional polar radius. The dash curve shows the same relation for the density. Note that both curves are close to unity for values of fractional polar radius less than 0.75. The curves are shown for the 8 $M_\odot$ model. The magnitudes of the departures from unity for both the temperature and density ratios decrease as the mass decreases.

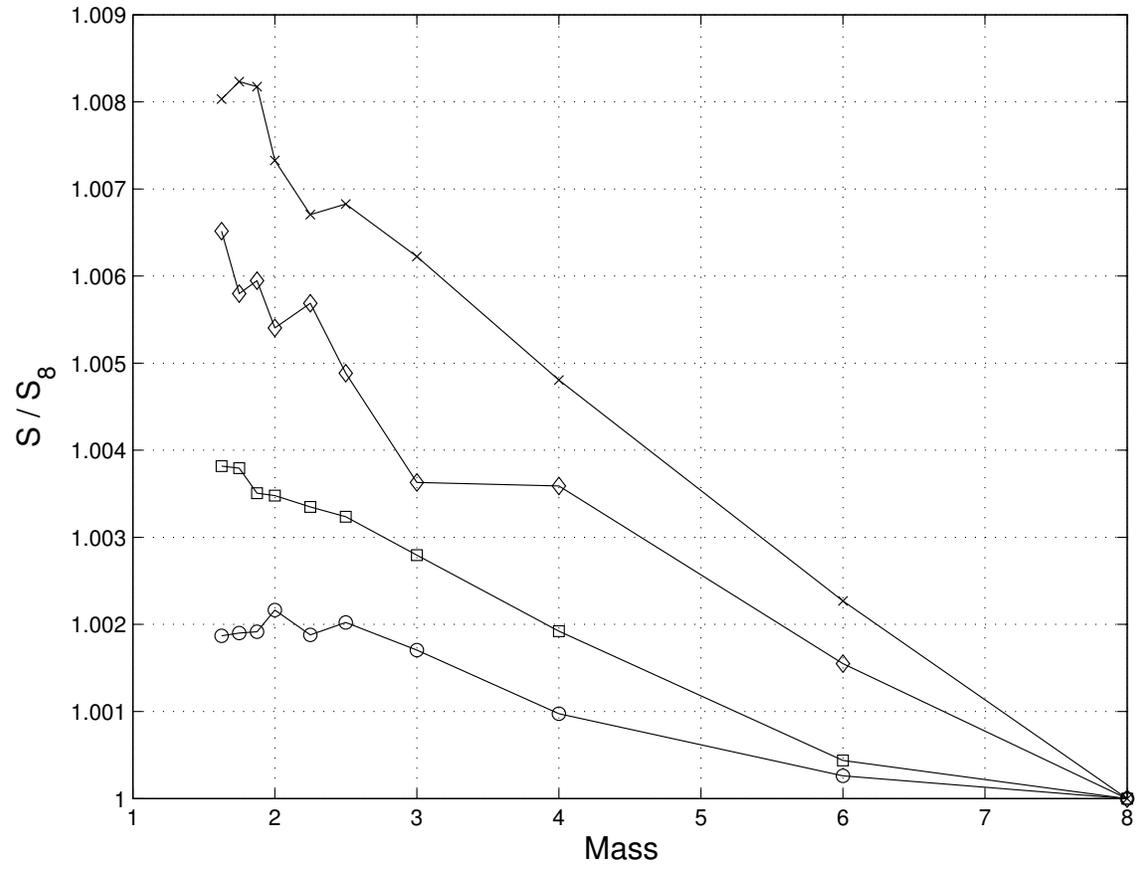

Fig. 1

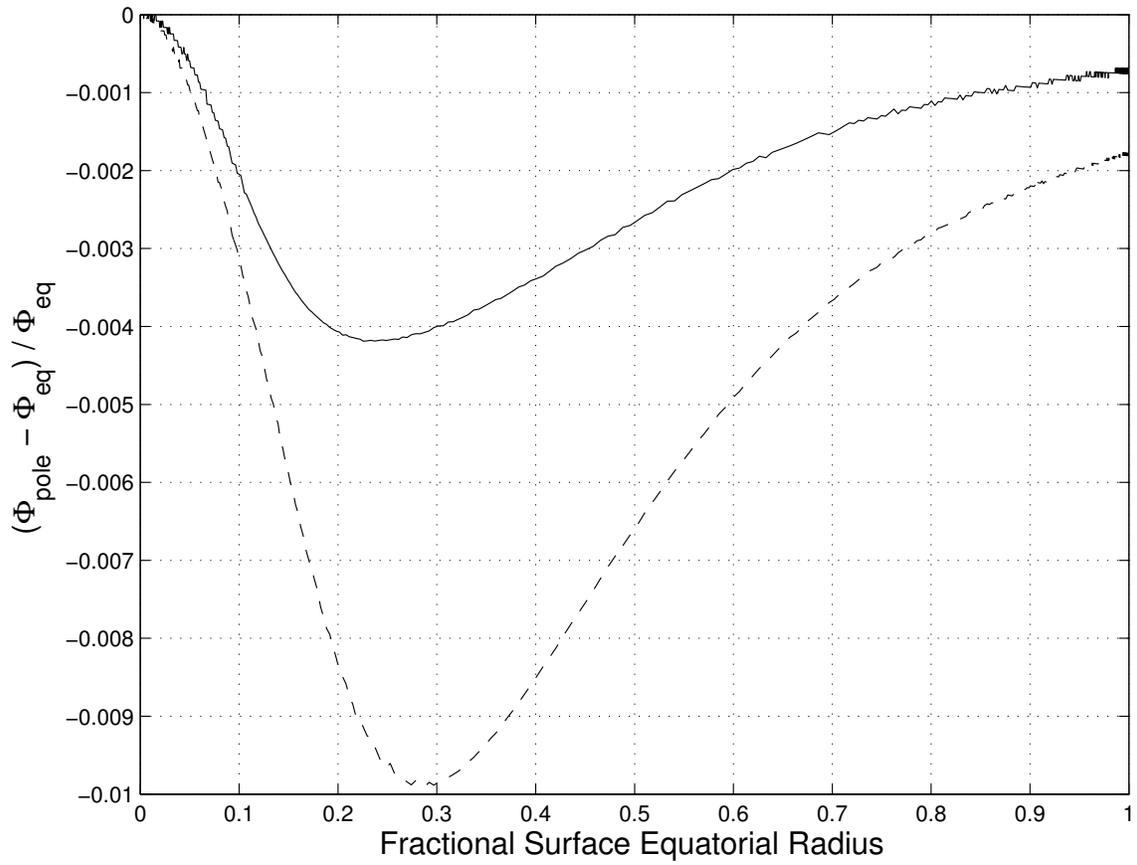

Fig. 2

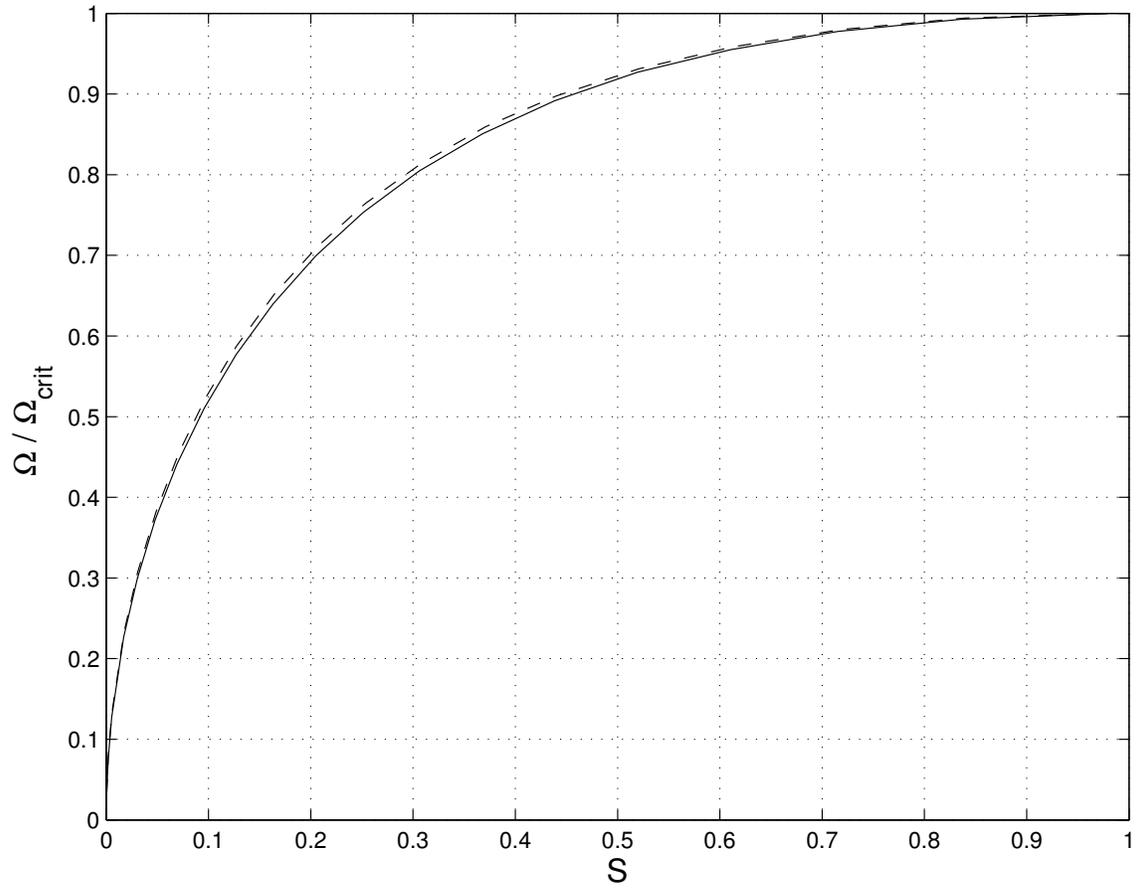

Fig. 3

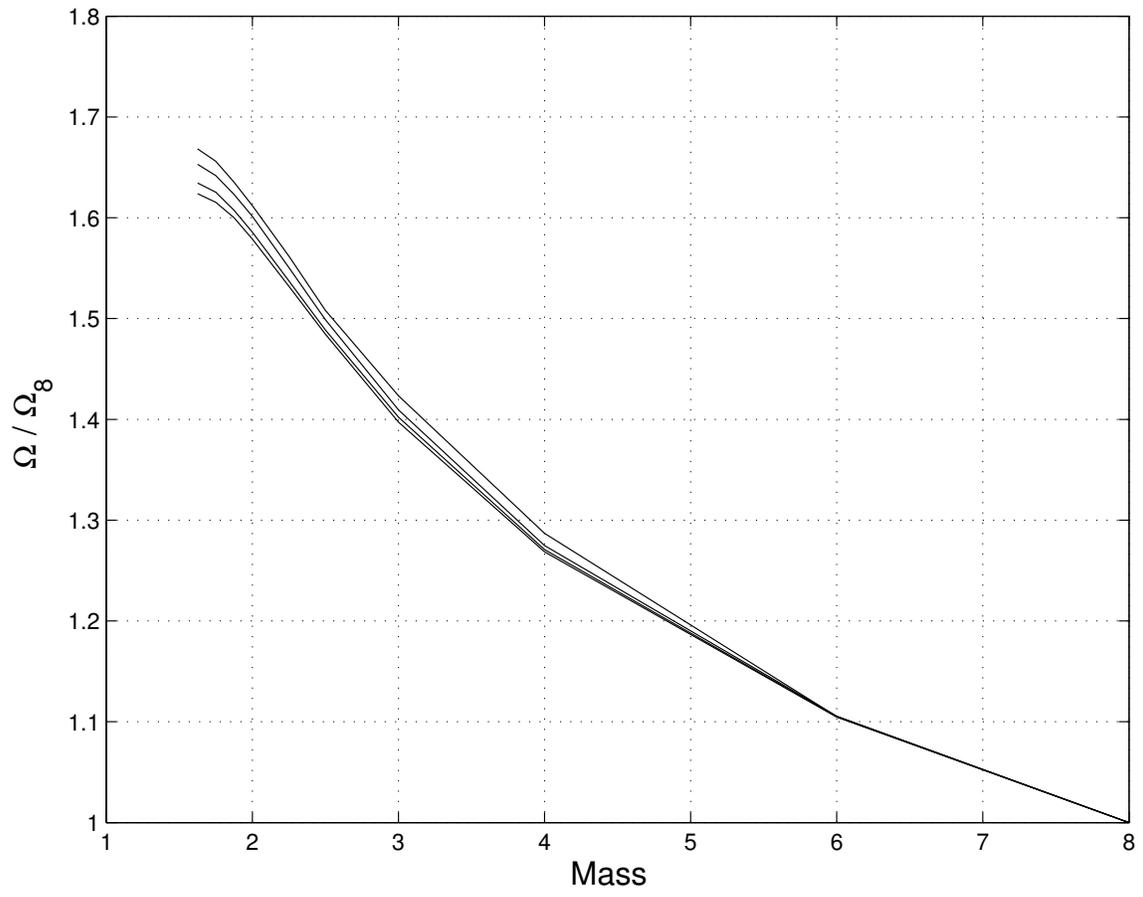

Fig. 4

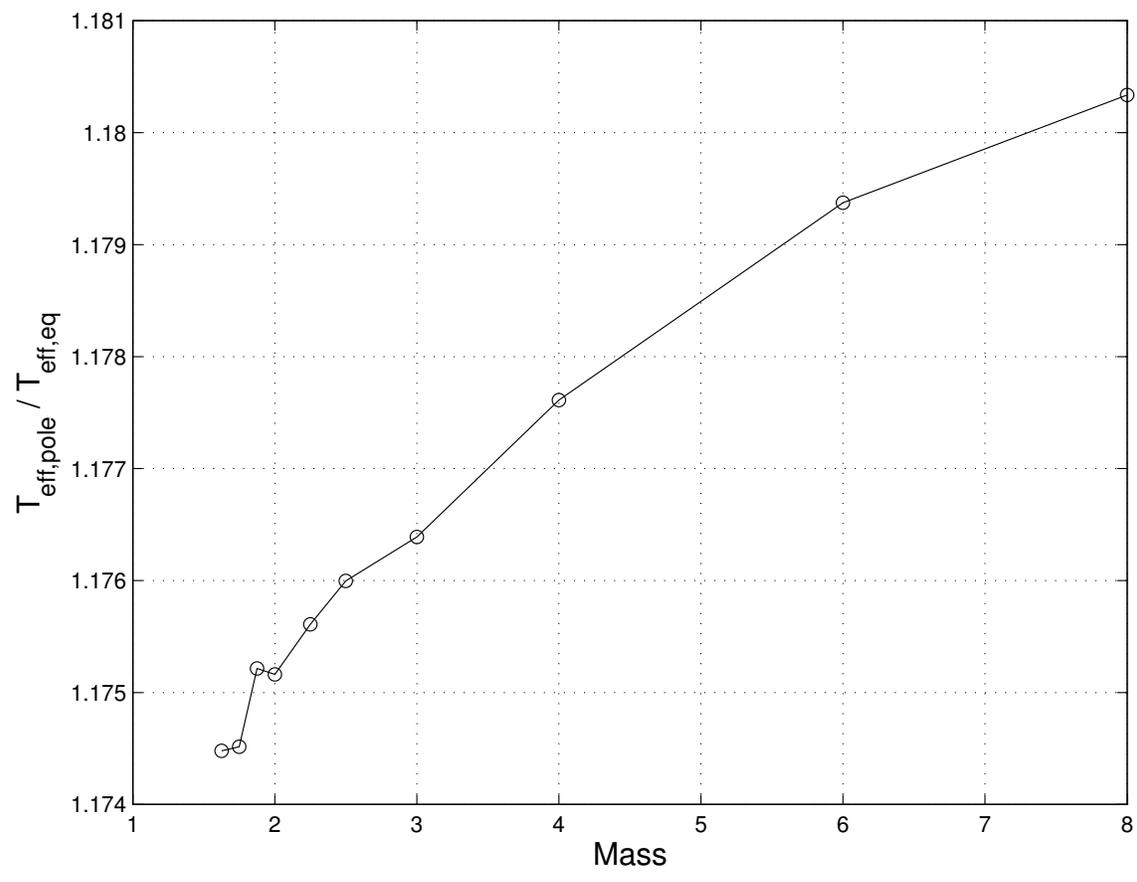

Fig. 5

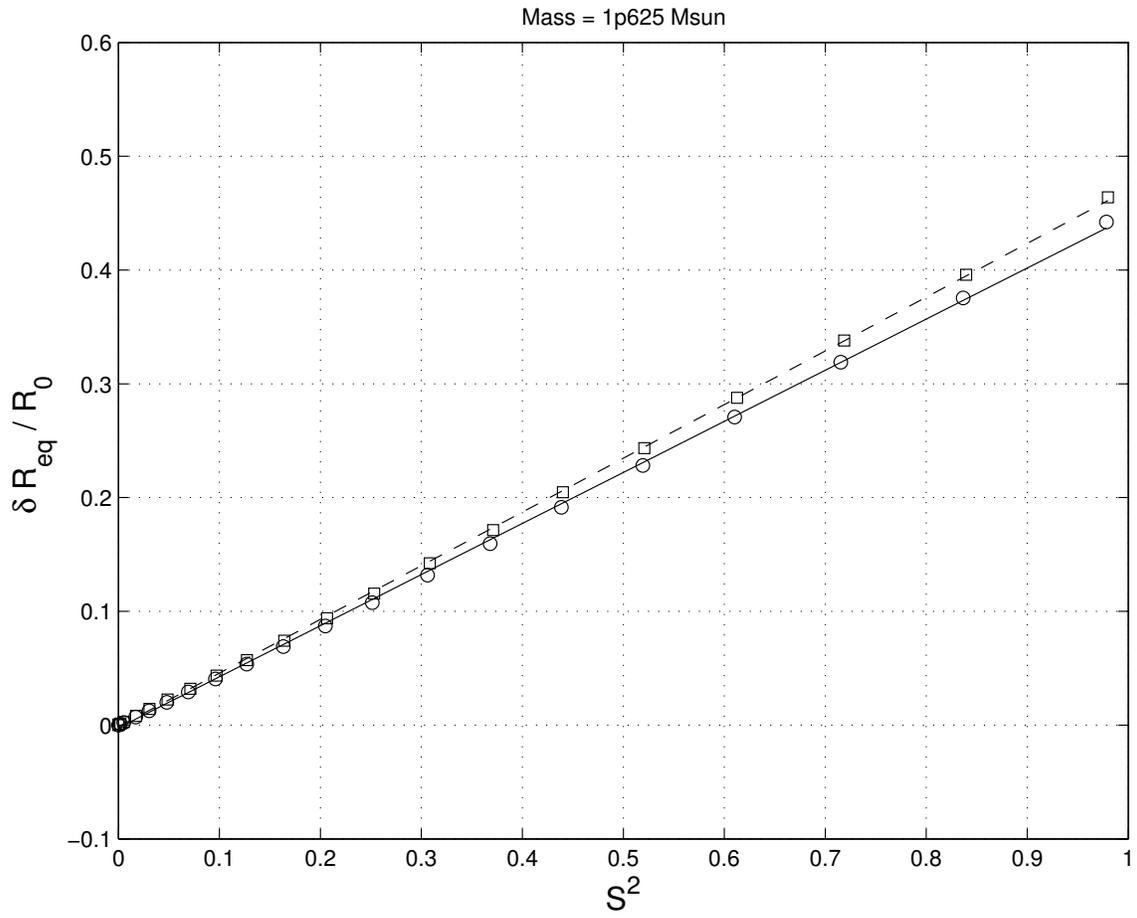

Fig. 6

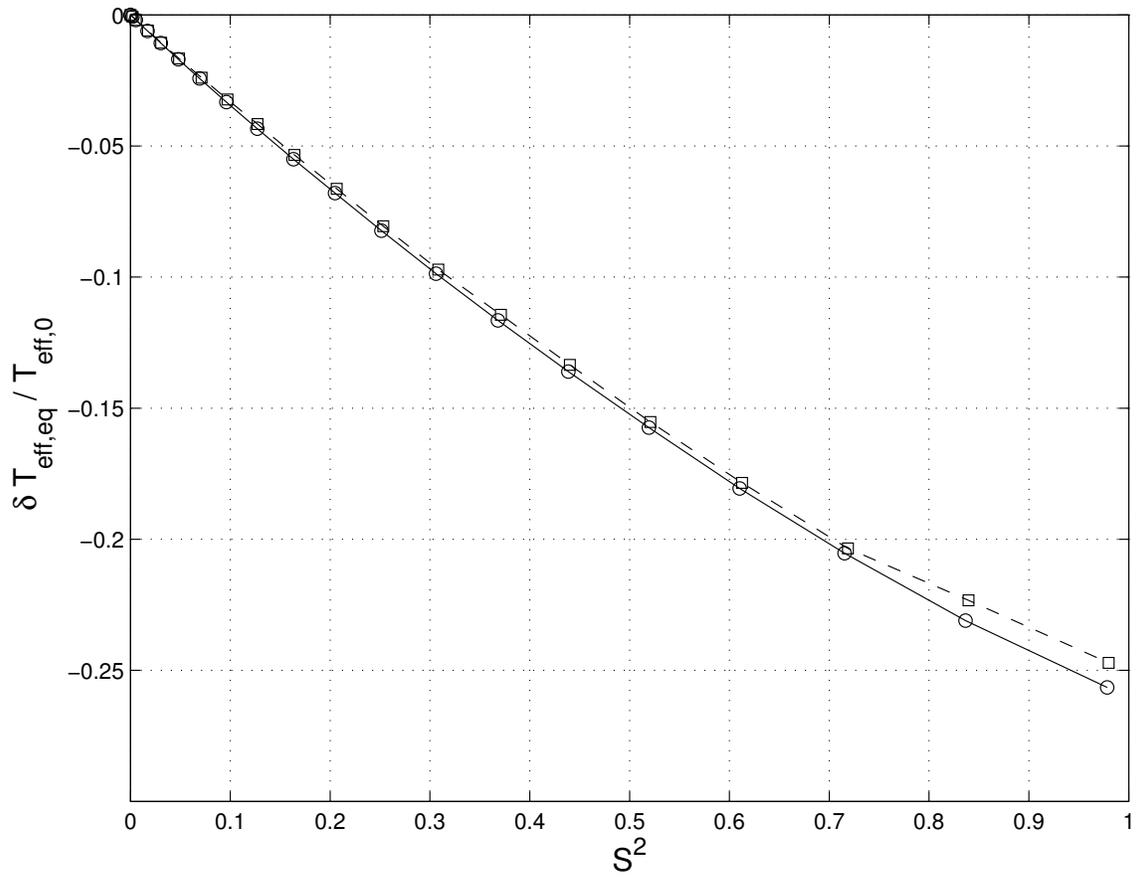

Fig. 7

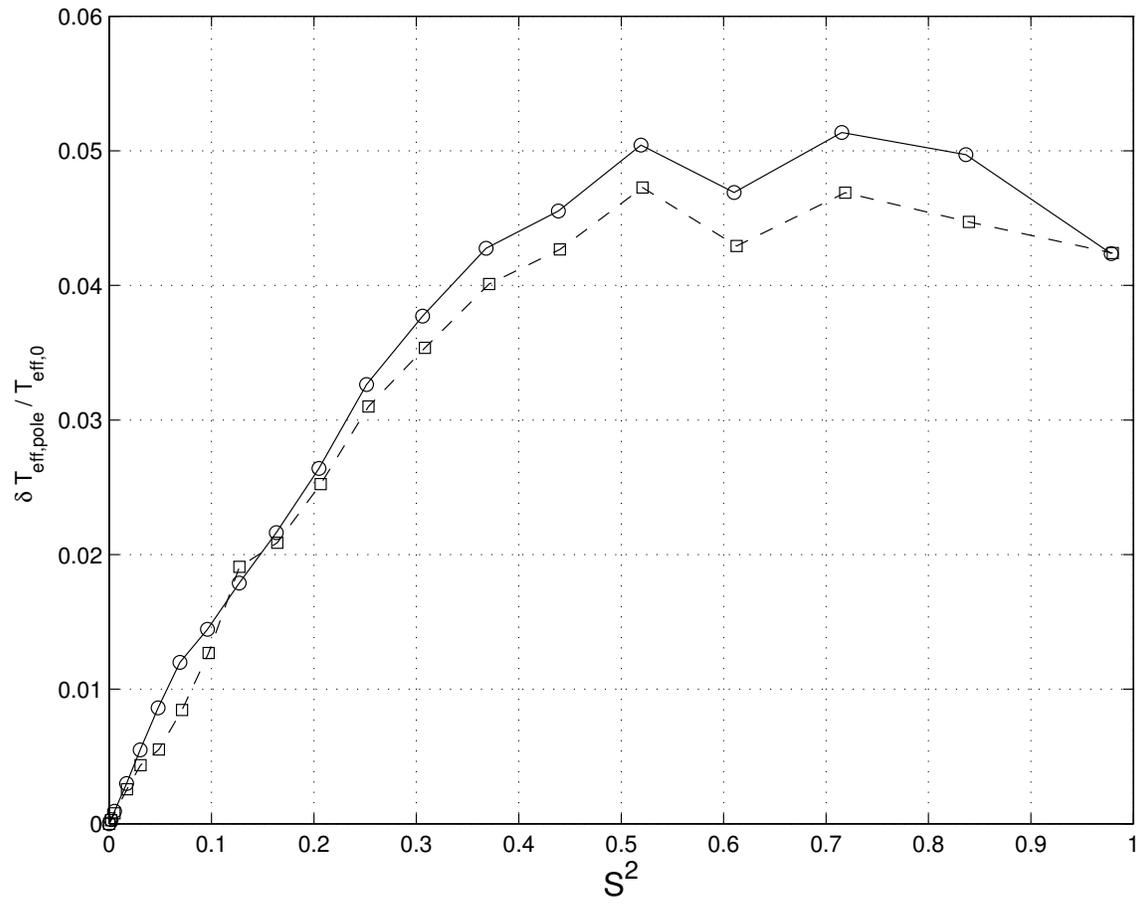

Fig. 8

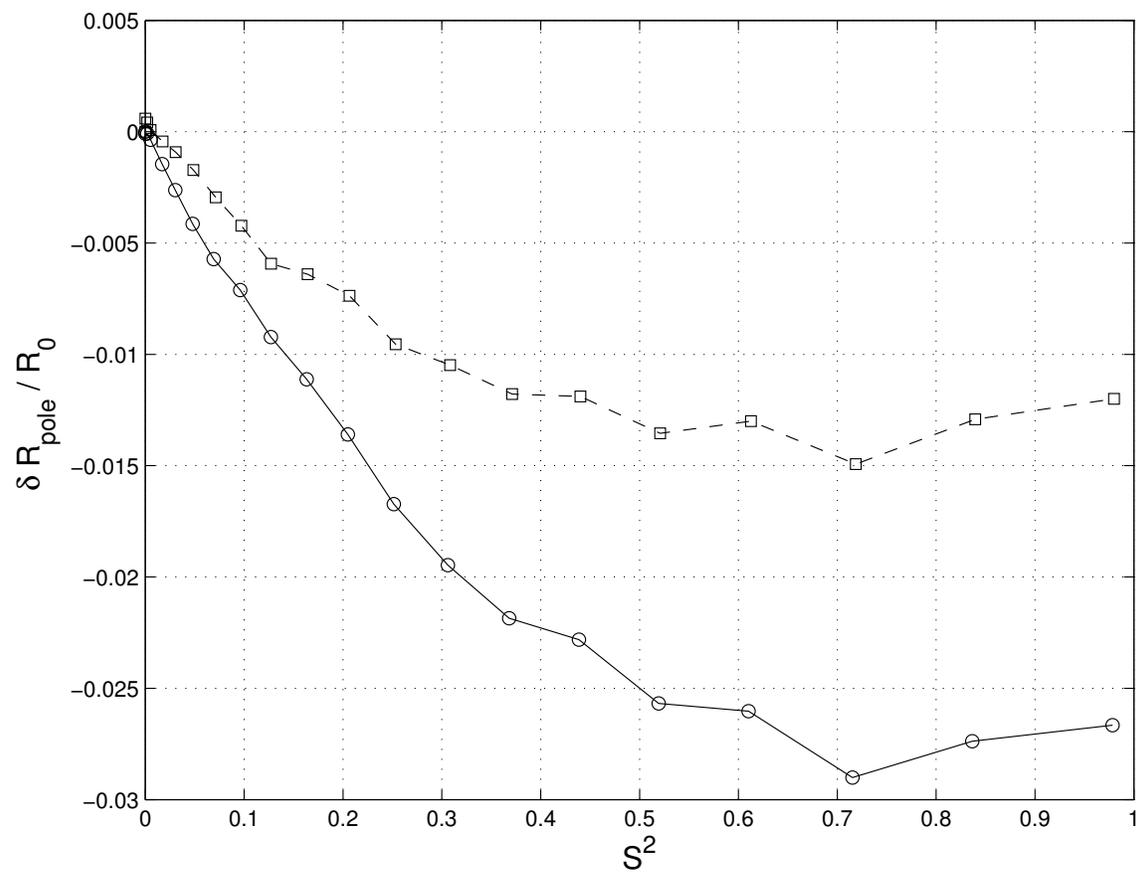

Fig. 9

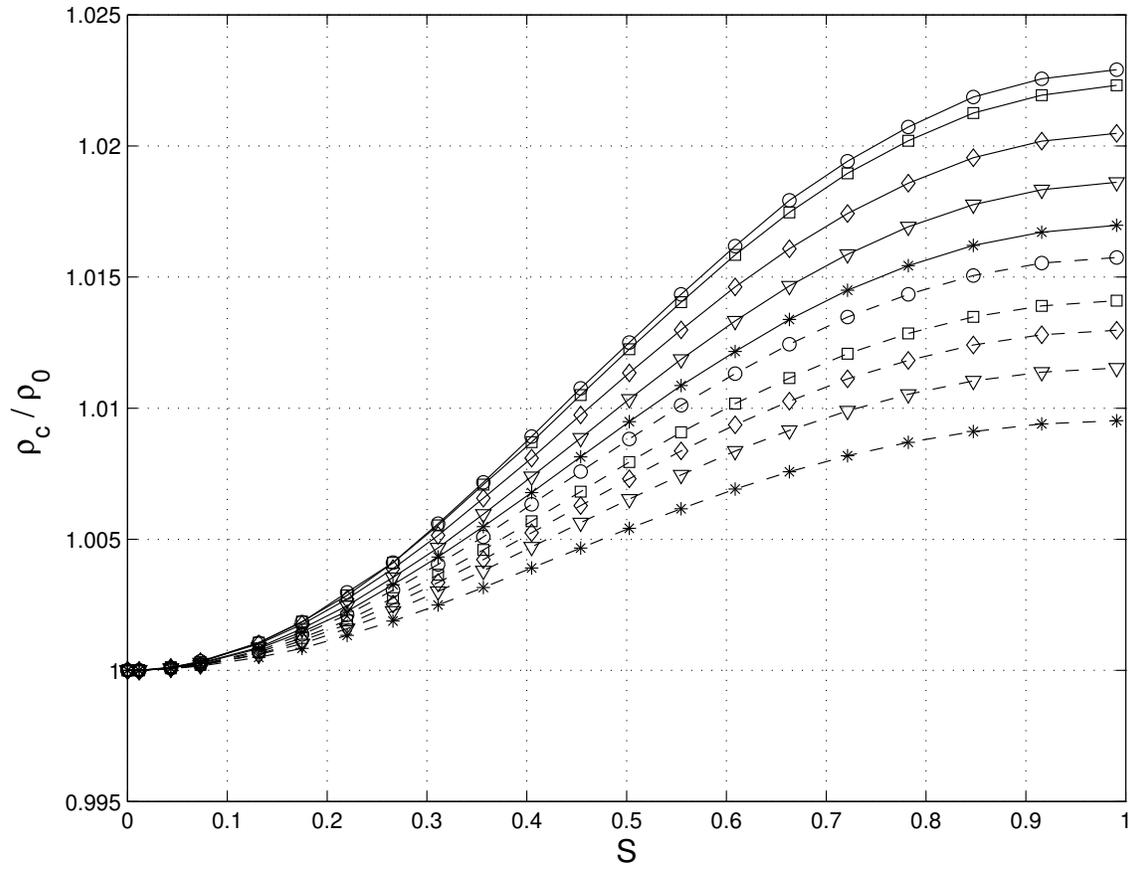

Fig. 10

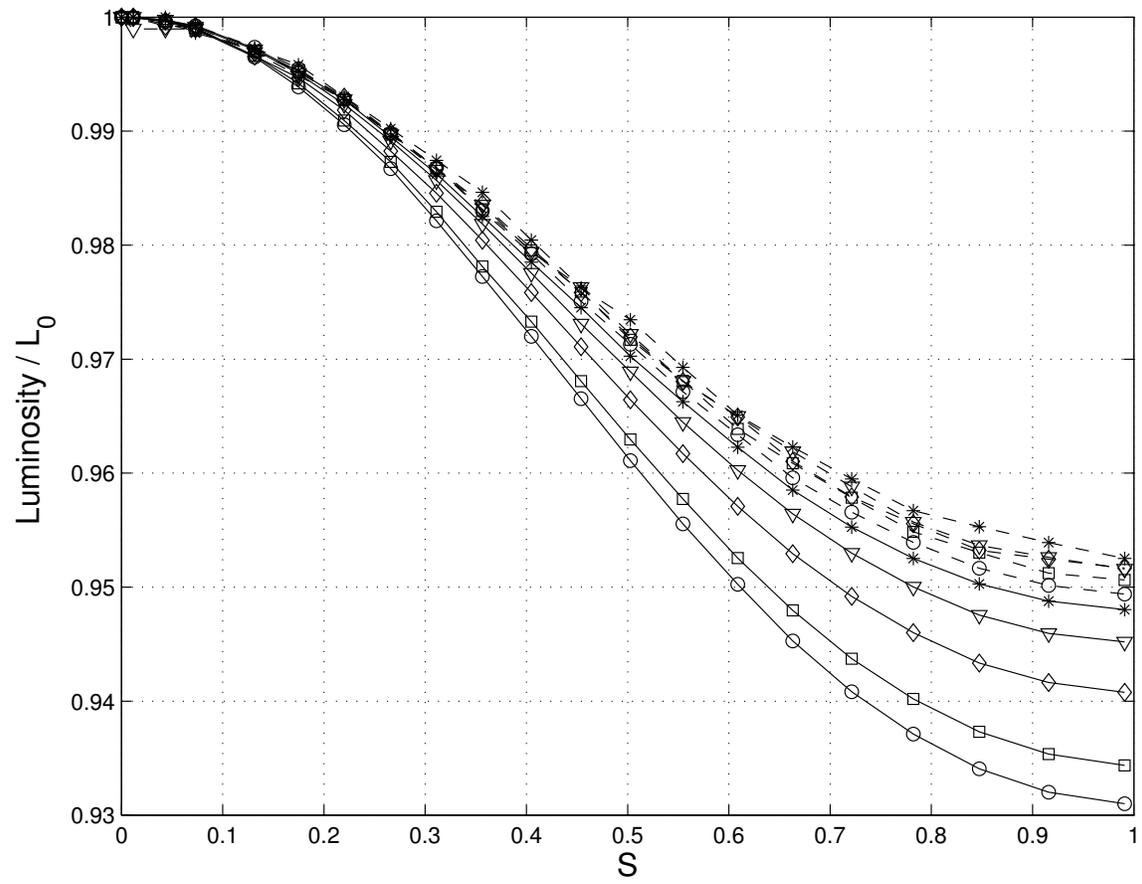

Fig. 11

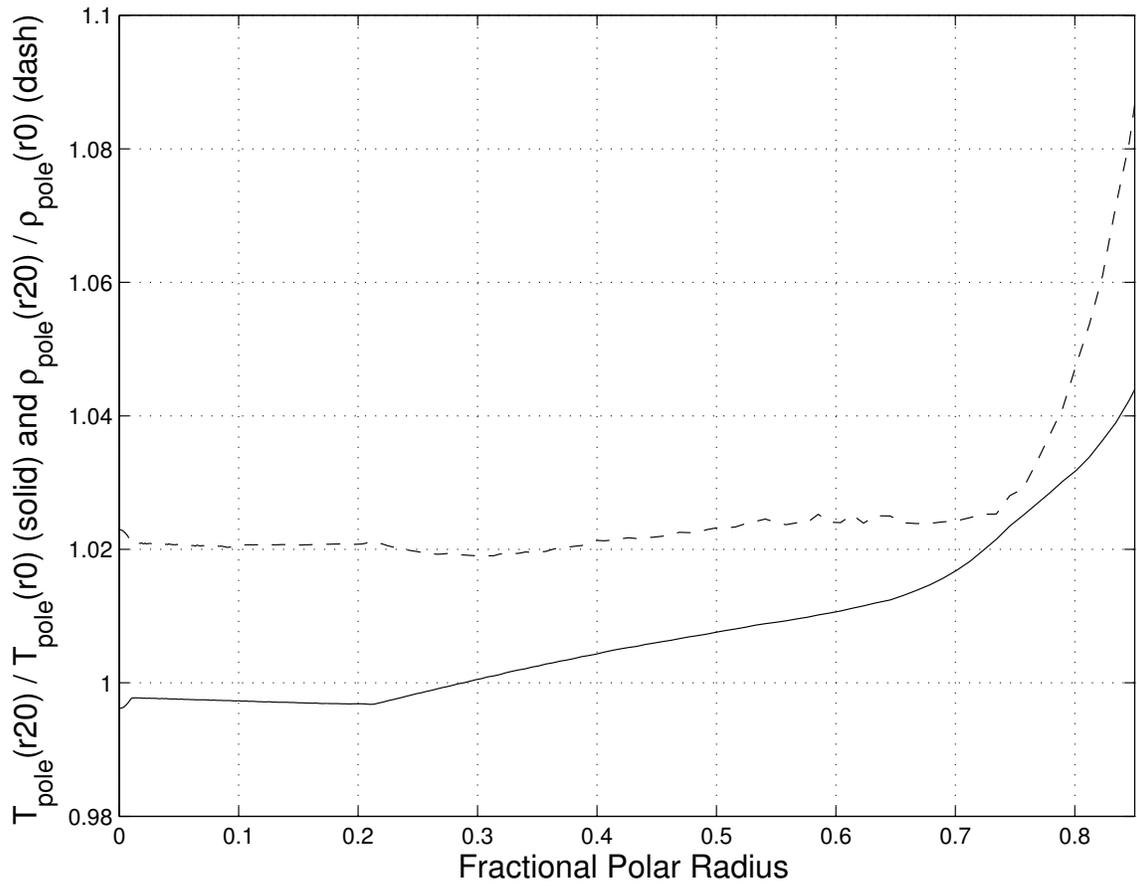

Fig. 12